\documentclass{aastex6}

\def\kms{km\,s$^{-1}$}
\def\Kkms{K\,km\,s$^{-1}$}
\def\dtwenty{d_{\rm 20}}

\def\schi{{\sc Hi}}
\def\msol{{$M_\odot$}}

\def\vlsr{v_{\rm LSR}}

\def\mchvc{M_{\rm CHVC}}

\def\simlt{\lower.5ex\hbox{$\; \buildrel < \over \sim \;$}}
\def\simgt{\lower.5ex\hbox{$\; \buildrel > \over \sim \;$}}
\def\schii{{\sc Hii}}

\fullcollaborationName{...}

\begin{document}


\title{A High-Velocity Cloud Impact Forming a Supershell in the Milky Way}


\author{Geumsook Park and Bon-Chul Koo} 
\affil{Department of Physics and Astronomy, Seoul National University \\
Seoul 151-747, Korea}

\author{Ji-hyun Kang}
\affil{Korea Astronomy and Space Science Institute \\
Daejeon 305-348, Korea}

\author{Steven J. Gibson}
\affil{Department of Physics and Astronomy, Western Kentucky University \\
1906 College Heights Blvd., Bowling Green, KY 42101, USA}

\author{J. E. G. Peek}
\affil{Space Telescope Science Institute \\
3700 San Martin Dr., Baltimore, MD 21218, USA}

\author{Kevin A. Douglas}
\affil{Department of Physics and Astronomy, Okanagan College \\
1000 K. L. O. Rd., Kelowna, British Columbia V1Y 4X8, Canada}

\author{Eric J. Korpela}
\affil{Space Sciences Laboratory, University of California \\
Berkeley, CA 94720, USA}

\and

\author{Carl E. Heiles}
\affil{Radio Astronomy Lab, UC Berkeley \\
601 Campbell Hall, Berkeley, CA 94720, USA}



\begin{abstract}
Neutral atomic hydrogen (\schi) gas in interstellar space is 
largely organized into filaments, loops, and shells, the  
most prominent of which are ``supershells''. 
These gigantic structures requiring $\ga~3\times~10^{52}$~erg to form
are generally thought to be produced by either 
the explosion of multiple supernovae (SNe) in OB associations or 
alternatively by the impact of high-velocity clouds (HVCs) 
falling to the Galactic disk. 
Here we report the detection of a kiloparsec (kpc)-size supershell in the 
outskirts of the Milky Way with the compact HVC 040+01$-$282 (hereafter CHVC040)
at its geometrical center
using the ``Inner-Galaxy Arecibo L-band Feed Array''
\schi\ 21-cm survey data.  
The morphological and physical properties of both objects suggest that 
CHVC040, which is either a fragment of a nearby disrupted galaxy 
or a cloud originated from an intergalactic accreting flow, 
collided with the disk $\sim 5$~Myrs ago to form the supershell.
Our result shows that some compact HVCs can survive their trip
through the Galactic halo 
and inject energy and momentum into the Milky Way disk.
\end{abstract}

\keywords{Galaxy: disk --- ISM: clouds --- radio lines: ISM}



\section{Introduction} \label{sec:intro}

Supershells are large gaseous shells of radius greater than
a few hundred parsecs.
They are distinct from other shell-like structures 
in their extraordinarily large energy requirement, i.e., 
$\ga 3\times 10^{52}$~erg, which corresponds to 
$\ga 30$ supernova (SN) explosions \citep{heiles1979,heiles1984,mcclure2002}.
About twenty supershells have been found in the Milky Way,
and numerous neutral atomic hydrogen (\schi) holes 
corresponding to supershells have been discovered 
in nearby dwarfs and spiral galaxies \citep{kamphuis1991,bagetakos2011}. 
These gigantic structures are generally thought to arise from 
multiple SN explosions in stellar OB associations.
But most supershells
are missing a stellar association in their interior, 
and the number of supershells and their energies are usually 
incompatible with the level of star formation in those galaxies 
\citep{heiles1984,rhode1999}.
Therefore, several alternative scenarios have been proposed, 
the most popular of which is the collision of high-velocity clouds (HVCs) 
with the disk \citep{tenorio1980,tenorio1987,mirabel1990}. 

HVCs are \schi\ clouds with radial velocities very different from 
the disk material in the Milky Way, 
e.g., with a deviation more than 50~\kms\
from the range of permitted velocities in a simple model of 
the distribution and rotation of the \schi\ gas in the Galaxy \citep{wakker1991a}.
Some large HVC complexes are known to 
be gas streams tidally stripped from satellite galaxies of the Milky Way, 
but the origin of isolated compact HVCs (CHVCs) remain controversial: they could be 
clouds formed from galactic fountain or intergalactic accreting flows,
part of the large HVC complexes, 
or condensations in the multi-phase circumgalactic medium \citep{putman2011, wakker2004, putman2012}.
The HVC origin has been proposed 
for a few supershells \citep{heiles1984, mirabel1990,tamanaha1997}, but 
there has been no clear example showing a direct link between the two, 
particularly for CHVCs. 

Here we report the detection of  
a Galactic supershell with an associated HVC, 
GS040.2+00.6$-$70 (hereafter GS040). 
GS040 was first identified as a faint, forbidden-velocity wing feature (FVW~40.0+0.5)
in the low-resolution, large-scale longitude-velocity study of \cite{kang2007}.
We have found that GS040 appears to be  
a complete circular ring with complicated structures inside 
in our high-resolution I-GALFA (Inner-Galaxy ALFA) 
\schi\ 21-cm line survey data. 
The I-GALFA survey is a survey 
of the first Galactic quadrant visible to Arecibo 
($\ell=32\arcdeg$ to 77\arcdeg\ and $|b| \la 15\arcdeg$) 
done by using the 7-beam Arecibo L-band Feed Array (ALFA) 
receiver on the Arecibo 305~m telescope, and it  
provides sensitive ($\Delta T_b=0.2$~K) and fully-sampled \schi\ maps
at spatial and spectral resolutions of 4\arcmin\ and 0.184~\kms, respectively \citep{koo2010,gibson2012}.
The I-GALFA survey data further reveal that there is 
a CHVC at the very center of GS040. This CHVC, named 
HVC 040+01$-$282 (hereafter CHVC040), was first identified in  
the Leiden/Dwingeloo survey \citep{wakker1991b} and was later classified as an isolated CHVC by \cite{braun1999} and \cite{deheij2002}.
\citet{westmeier2005} presented 
a higher-resolution (9\arcmin) \schi\ image obtained from  
the Effelsberg telescope, which showed that CHVC040 has a 
pronounced head-tail structure. 
Our Arecibo \schi\ images reveal  
detailed spatial and velocity structure of CHVC040 
strongly suggesting its association with the supershell GS040.
We describe two structures in Section~\ref{sec:targets},
and discuss their physical characteristics 
and their association together with some implications on 
the disruption of HVCs in Section~\ref{sec:disc}.

\section{Supershell GS040.2+00.6$-$70 and High-Velocity Cloud HVC040+01$-$282}
\label{sec:targets}

\begin{figure}[t]
\begin{center}
\epsscale{1.2}
\plotone{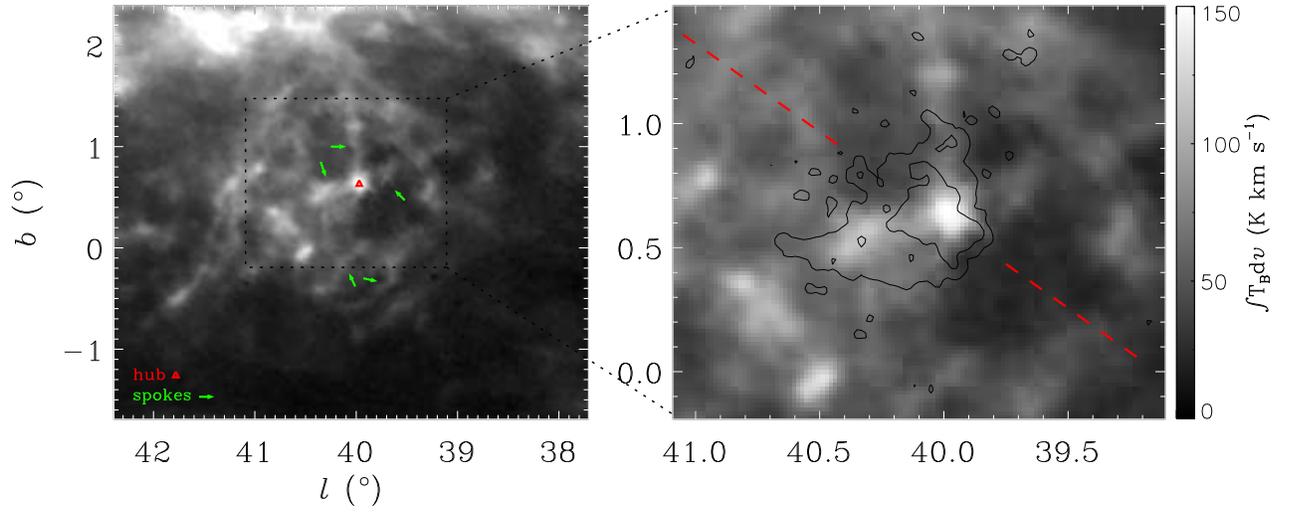}
\caption{
\schi\ integrated-intensity maps of 
the overall picture of the supershell GS040 (left) 
and a close-up view inside its ring structure (right).
The maps are obtained by integrating the emission between 
$\vlsr=-141$ and $-66$~\kms.
The features described in Sec.~\ref{sec:targets} are
labeled in the left image.
The overlaid contours in the right image show the overall appearance of the high-velocity cloud CHVC040
in integrated \schi\ emission (see Figure~\ref{fig:hvc}).
The position-velocity diagram in Figure~\ref{fig:pvmap} was obtained along the dashed red line.
\label{fig:itg_gs040}}
\end{center}
\end{figure}

\begin{figure}[t]
\begin{center}
\epsscale{1.2}
\plotone{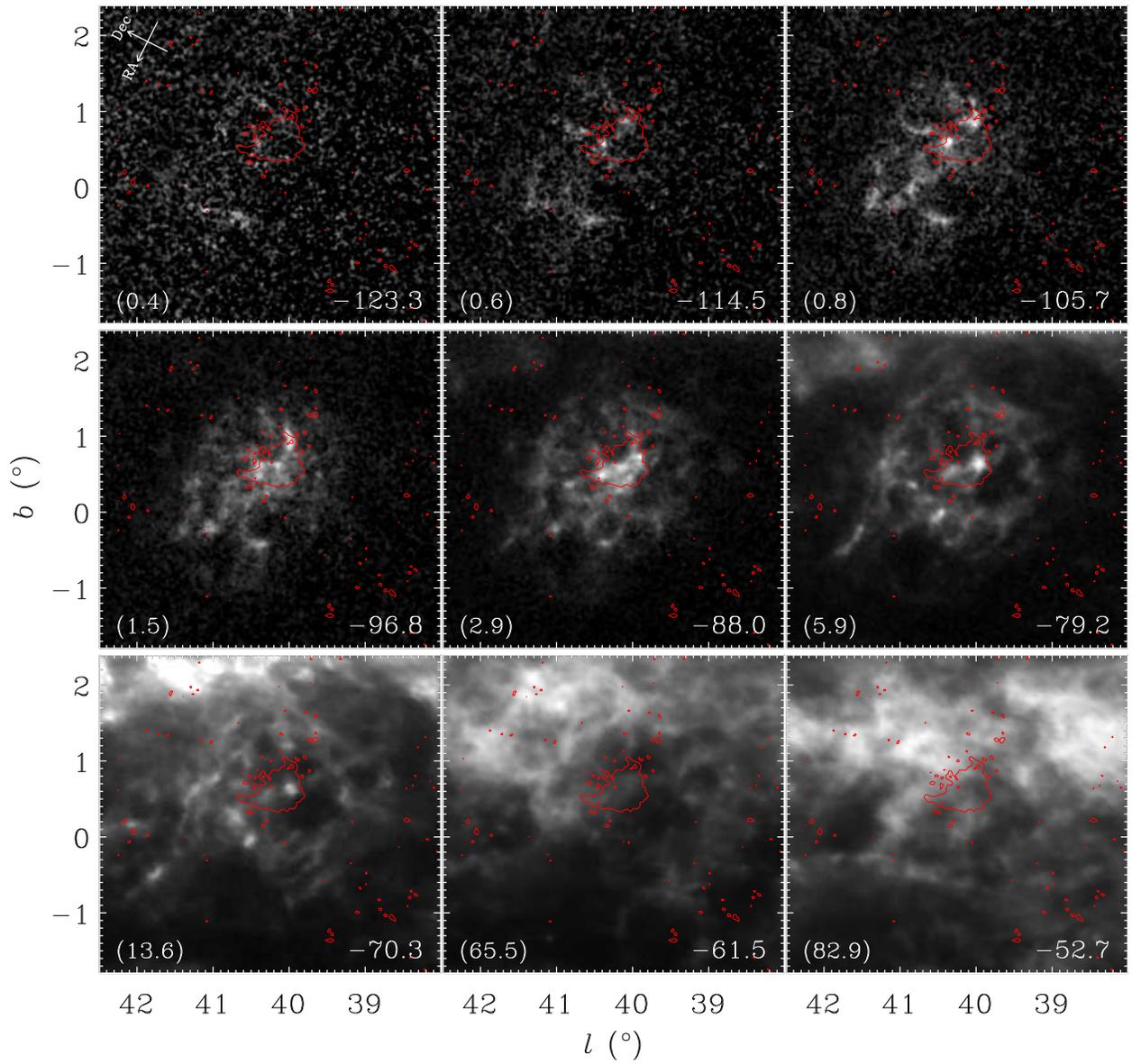}
\caption{
Velocity-channel maps of GS040 in the \schi\ data.
The central LSR velocity of each channel is written
at the bottom right corner in \kms.
The channel width and interval are 3.68 and 5.15~\kms, respectively.
The \schi\ brightness temperature in gray scale ranges
from 0~K (black) to the value (white) in parentheses
at the bottom left corner of each panel in Kelvins.
Contours show the overall appearance of CHVC040
in integrated \schi\ emission (see Figure~\ref{fig:hvc}).
To aid comparison to literature studies in Equatorial coordinates,
arrows are shown in the first panel indicating Right Ascension and Declination
(J2000) coordinate directions.
\label{fig:chmap_gs040}}
\end{center}
\end{figure}

The supershell GS040 is centered at $(\ell,b) = (40\fdg2, +0\fdg6)$ and 
clearly visible from $\vlsr \sim -120$ to $-70$~\kms.
In an integrated intensity map (Figure \ref{fig:itg_gs040}), 
GS040 appears as a complete circular ring of radius $\sim1\fdg3$ 
with complicated structures inside.
More detailed structures can be seen 
in Figure~\ref{fig:chmap_gs040}, which presents velocity channel maps.  
At the most negative velocities ($\sim -120$~\kms), we see 
diffuse emission with embedded knotty filaments near the center and 
an extended filament in the south\footnote{Directions in this paper
are all in reference to Galactic coordinates, not J2000 Equatorial coordinates.}.
The features are observable at even 
more negative velocities ($-150$~\kms~$< \vlsr < -120$~\kms)
but they are extremely weak and hard to detect in velocity-channel maps.
As the velocity increases ($\vlsr\ga -90$~\kms),
the nebulosity fades out, and a larger ($\sim 2\fdg6$) ring structure appears,  
which resembles a cartwheel 
with a bright central ``hub'' and several ``spokes'' (see also Figure~\ref{fig:itg_gs040}). 
The size of the ring increases slightly with velocity 
indicating that the ring structure is an approaching portion of an expanding shell. 
At velocities greater than about $-70$~\kms, 
the Galactic background \schi\ emission becomes dominant,
and the emission associated with GS040 is less clear. 

At the very center of GS040 is the HVC CHVC040. 
The positional coincidence of CHVC040 with the GS040's central hub  
is striking as can be seen in Figure~\ref{fig:pvmap}, which is the 
position-velocity map crossing the center of GS040.
The morphological agreement between the two is also noticeable. 
Our high-resolution Arecibo \schi\ image reveals that 
CHVC040 has a blunt cone shape with a steep southwestern boundary 
and a faint envelope flaring out northeast (Figure~\ref{fig:hvc}).
This morphology of CHVC040 matches well 
with that of the central hub of GS040, e.g., see Figure~\ref{fig:itg_gs040} 
and also the channel map 
at $-88$~\kms\ in Figure~\ref{fig:chmap_gs040}. 

Figure~\ref{fig:hvc} shows the detailed spatial and velocity structures of CHVC040.
The integrated intensity map in the left frame shows that 
CHVC040 has a bright, $\sim 12' \times 15'$- sized ``core'' 
elongated along the northeast-southwest direction.
The core appears to be composed of several clumps 
with a sharp boundary at southwest, 
whereas the diffuse envelope appears to be slightly more extended toward southeast. 
The mean \schi\ column density, assuming that the emission is optically thin, 
is 1.5$\times~10^{19}\,{\rm cm}^{-2}$ while the peak \schi\ column density is 
about two times higher. 
The velocity centroid map in the middle frame shows that 
systemic velocity attains its most negative ($\simlt -290$~\kms) at the southwestern boundary 
and most positive  ($\sim -260$~\kms) along the southeastern boundary 
and also in the middle of the northwestern boundary. 
The mean systematic velocity is $-282$~\kms. 
If the most positive feature in the northwestern boundary were not present, the 
velocity structure could have been suggestive of a rotation   
with respect to the northeast-southwest  
symmetry axis at speed of $\simlt 20$~\kms. 
The velocity width (Full Width at Half Maximum; FWHM) 
ranges 25--45~\kms\ over most parts of the cloud with a median of 36~\kms. 
The southwesternmost thin layer has relatively narrow width ($\sim$19--28~\kms) 
while parts of the northern area has width $\ga 50$~\kms. 
The region with the narrowest line width is very thin ($\sim 3\arcmin$),
and therefore \citet{westmeier2005} could not spatially resolve it.

Some representative line profiles are shown in Figure~\ref{fig:hvc_prf} together 
with the average line profile of the cloud. 
Most profiles are well described with 
a single Gaussian component, while 
some profiles, e.g., c and e-g, show clear double peaks or a narrow 
component superposed on a broad component. 
The velocity width of the mean profile is 
45~\kms, which is considerably larger than 
the median value (36~\kms) of individual profiles,
presumably due to the dispersion of central velocities.
The median width corresponds to 
a kinetic temperature of $2.8\times 10^4$~K.
At several positions, narrow velocity components are detected, 
but still their widths are $\ge 10$~\kms, which
implies a kinetic temperature $\ge 2,\!000$~K.
This indicates that CHVC040 is mostly composed of warm neutral gas, 
which is not unusual for CHVCs \citep{winkel2011,faridani2014}.
But it is different from those 
head-tail HVCs with a narrow-line head of undisturbed cold \schi\ gas 
and a wide-line tail of disturbed warm \schi\ gas 
\citep[e.g., HVC125+41-207;][]{bruns2001}.

\section{Discussion}
\label{sec:disc}

\subsection{Formation of GS040 by the Collision of CHVC040}
\label{sec:disc_sub1}

\begin{figure}[t]
\begin{center}
\epsscale{.8}
\plotone{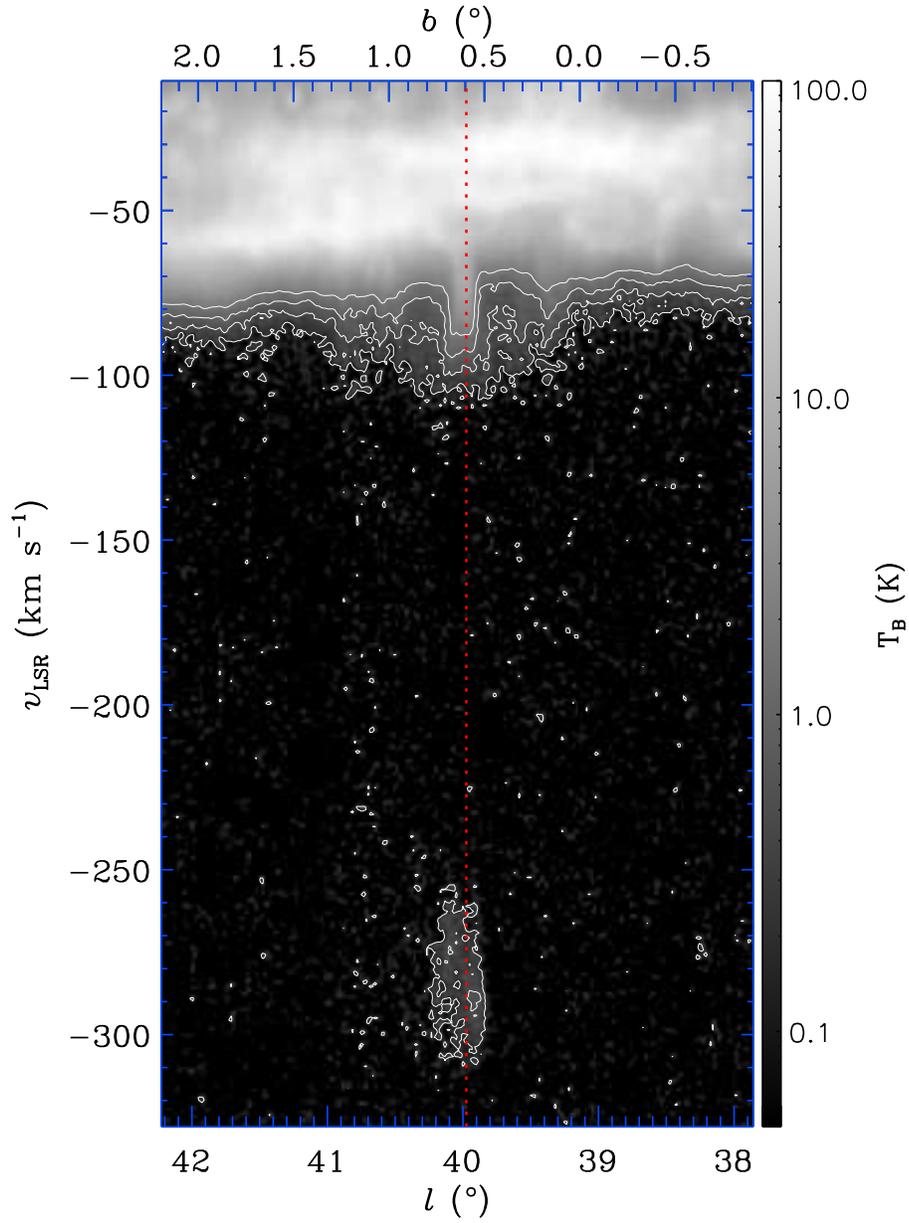}
\caption{
Position-velocity map of \schi\ emission of GS040 and CHVC040. 
This diagram is drawn along a line crossing the center of GS040 at 
position angle of about 35\arcdeg\ (Figure~\ref{fig:itg_gs040}). 
The lower and upper x-axis indicate 
$\ell$ and $b$ coordinates of the position, respectively.
The y-axis indicates the LSR velocity. 
The contour levels are 0.2, 0.5, 1, and 2~K in brightness temperature.
The dotted vertical line is toward the center of GS040.
\label{fig:pvmap}}
\end{center}
\end{figure}

\begin{figure}[t]
\begin{center}
\epsscale{1.2}
\plotone{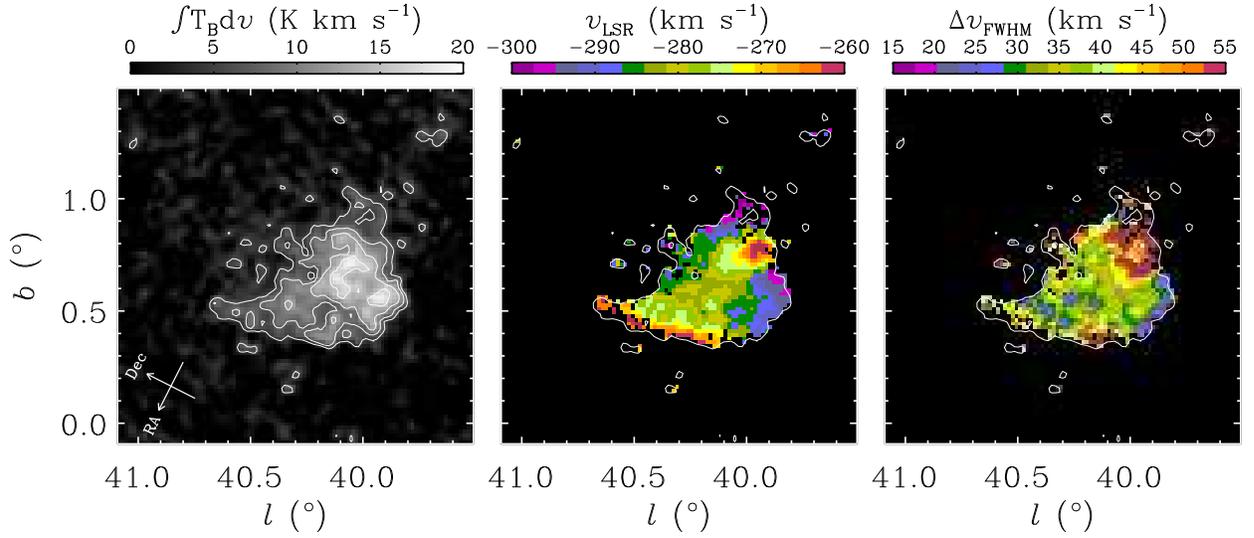}
\caption{
Spatial and velocity structures 
of the high-velocity cloud CHVC040.  
{\it Left:} \schi\ intensity map integrated over LSR 
velocities from $-330$ to $-230$~\kms. 
Contours are drawn at 5, 8, 12, and 16~\Kkms.
If the emission is optically thin, 1~\Kkms\ corresponds to 
hydrogen columns of $1.82\times 10^{18}$~cm$^{-2}$.
{\it Middle and Right:} 
Central velocity and velocity width maps 
derived by fitting single Gaussian curve to each line profile.   
The fit is limited to positions with integrated intensity 
greater than 5~\Kkms, i.e., within the white contour.
The pixels where the emission is weak and a reasonable fit cannot be obtained 
are also blanked out.
To aid comparison to literature studies in Equatorial coordinates,
arrows are shown in the left panel indicating Right Ascension and Declination 
(J2000) coordinate directions.
\label{fig:hvc}}
\end{center}
\end{figure}

\begin{figure}[t]
\begin{center}
\epsscale{1}
\plotone{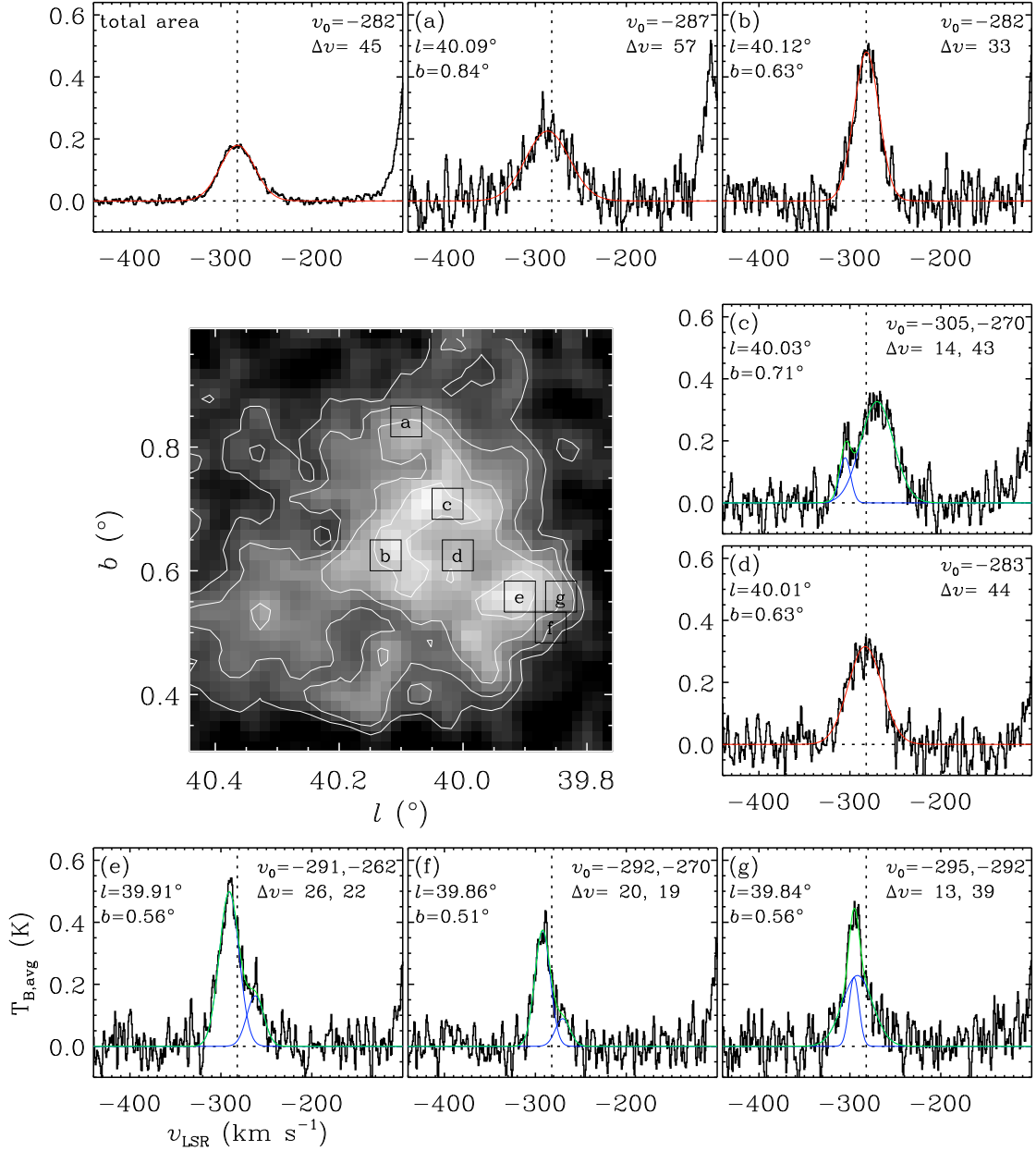}
\caption{
Examples of CHVC040 \schi\ line profiles.
The profile in the top left frame is an average profile of the whole cloud.
The rest of line profiles are extracted from the areas marked by boxes
in the central image, which is the integrated intensity map in Figure~\ref{fig:hvc}.
The profiles have been Hanning smoothed 
to have a velocity resolution of 2.96~\kms.
A Gaussian fit with the suitable number (n) of components is applied
and displayed with a red (n=1) or
green curve (n=2; each in a blue line).
Resultant parameters, including
central LSR velocity ($v_0$) and velocity width ($\Delta v$) in \kms,
are written on the top right corner of each panel.
A vertical line at the central velocity of the average line profile ($-282$~\kms)
is drawn to aid the intercomparison of different plots.
\label{fig:hvc_prf}}
\end{center}
\end{figure}

The location of CHVC040 at the geometrical center of GS040 suggests
that their physical association is very likely. 
The centroids of the CHVC emission 
and that of the GS040's hub emission 
overlap within $\sim 0.2$~degrees.  
The probability of this being a random alignment is 
$\sim 3 \times 10^{-6}$, and multiplying this by $\sim 300$ CHVCs 
yields an overall probability of $9\times 10^{-4}$ 
for any CHVC aligning this well with the GS040's hub. 
No intermediate-velocity \schi\ connecting the two is apparent 
(Figure~\ref{fig:pvmap}), but this could be because the gas is ionized.
We searched for a warm ionized gas 
associated with GS040 using 
the Wisconsin H$\alpha$ Mapper Northern Sky Survey (WHAM-NSS) data 
\citep{haffner2003}. The survey has an angular resolution of 1\arcdeg, and,    
around the GS040 area, it provides H$\alpha$ spectra 
covering $\vlsr$ from $-85$ to $+100$~\kms\ at  
spectral resolutions of 12~\kms. 
We have examined the H$\alpha$ intensity map 
integrated over the velocity range of GS040 ($-85$ to $-66$~\kms),
but could not detect an associated emission 
($\simlt$ 0.05~$R$ where 
$1 R = 10^6/4\pi\,{\rm photons\,cm}^{-2}\,{\rm sr}^{-1}\,{\rm s}^{-1}$).
Note that this area is bright in the total H$\alpha$ intensity map 
with a mean intensity of about 3~$R$,  
so that we do not expect to see the faint 
emission associated with either 
GS040 or CHVC040 in the all-sky H$\alpha$ maps 
\citep{finkbeiner2003,dennison1998}. 
We also searched for an associated hot ionized gas using the 0.1-2.4 keV image of 
the ROSAT All-Sky X-ray Survey \citep[1 pixel scale = 44\arcsec;][]{voges1999},
but couldn't detect any emission.

CHVC040 belongs to the ``Galactic Center Negative'' (GCN) HVC complex, which is 
a collection of small discrete HVCs sparsely distributed over 
a $70\arcdeg\times 70\arcdeg$ area 
within $\ell=0\arcdeg$ to $70\arcdeg$ and 
$b=-60\arcdeg$ to 10$\arcdeg$ \citep{wakker1991b,winkel2011}. 
A kinematic model has been proposed where 
GCN is a smooth gas flow starting  at $b=-60\arcdeg$ 
at a heliocentric distance of 35~kpc 
and crossing the Galactic plane obliquely at 15~kpc \citep{jin2010}.
For comparison, GS040 is probably located near the 
Scutum-Centaurus (Sct-Cen) arm at a distance of $\sim 20$~kpc \citep{dame2011}
because the disrupted interstellar medium (ISM)
is seen only at velocities below that of the  
Sct-Cen arm ($\sim -60$~\kms), not at higher velocities (see Figures~\ref{fig:chmap_gs040} and \ref{fig:pvmap}).
We examined lists of known stellar objects,
\schii\ regions, OB stars, and SN remnants (SNRs), but
no known sources are likely to be associated with GS040.
There is one \schii\ region (G$039.864+00.645$) in the {\it WISE} catalog of
Galactic \schii\ regions \citep{anderson2014} 
that is located at $\sim 6\arcmin$ west of the hub.
But this \schii\ region has a systematic velocity of
$-40.9$~\kms\ \citep{anderson2011} and is enclosed by a small (68\arcsec)
dust shell, so it cannot be responsible for the HI shell.
Instead the distance of $\sim 20$~kpc is not unreasonable for CHVC040,  
because GCN does not have a smooth extended envelope 
like other HVC complexes, and it appears to be composed of several 
subpopulations that do not share a common origin \citep{winkel2011}. 
Note that the geometrical center of GS040 is well below the Galactic plane 
($\sim 420\,\dtwenty$~pc where $\dtwenty\equiv d/20~{\rm kpc}$), 
i.e., at $b=+0\fdg6$ while the midplane there is at $b \sim +1\fdg8$
because the Galactic plane is warped in the outer Galaxy \citep{levine2006}. 
It is difficult to imagine the SN origin for a supershell at such height, and
CHVC040 is most likely the energy/momentum source for GS040. 

The total energy deposited ($E_E$) in the Galactic disk by CHVC040
can be inferred from 
the parameters of the GS040 supershell. 
The radius of GS040 is $450\,\dtwenty$~pc, while 
its mass at $\vlsr \le -75$~\kms\
is $1.6\times 10^5\,\dtwenty^2$~\msol\ 
including the cosmic abundance of helium. 
If we account for the mass unobservable due to Galactic background 
\schi\ emission, the total mass of GS040 
would be considerably greater. 
Adopting $v_s\sim 30$~\kms\ as the expansion speed of the shell, 
its kinetic energy is $E_K\ga 1.4\times 10^{51}\,\dtwenty^2$~erg. 
The collision should have occurred  
$\sim (1/3) R_s/v_s \sim 5\times 10^6\,\dtwenty$~yr ago, 
where the numerical factor $1/3$ accounts for the deceleration of the shell. 
Note that $E_K$ is a small fraction of the total energy deposited ($E_E$), 
most of which should have been radiated away. 
If GS040 was produced by multiple SNe, then, 
assuming instantaneous energy injection \citep{heiles1979}, 
$E_E\sim 5.3\times 10^{43}\, n_0^{1.12}\, R_s^{3.12}\, v_s^{1.4}
\sim 1.2\times 10^{53}\, (n_0/0.1~{\rm cm}^{-3})$~erg,  
where $n_0$ is ambient hydrogen density. 
At the position of GS040, the mean \schi\ density in the midplane is $\sim 0.1\,{\rm cm}^{-3}$,
and the \schi\ scale height is about 720~pc \citep{levine2006}.
So $n_0\sim 0.06~{\rm cm}^{-3}$, and we have $E_E\sim 7\times 10^{52}$~erg.   
For comparison, the 
total extent of CHVC040 is $210\,\dtwenty\times 320\,\dtwenty\,{\rm pc}^2$ 
while its \schi\ mass is $\mchvc=5,\!800\,\dtwenty^2$~\msol.
The area used to derive this mass has a geometrical mean radius of 150~pc, 
so that the mean hydrogen density of CHVC040 is 0.017~cm$^{-3}$. 
If CHVC040 collided with the rotating disk with the
mean `deviation' speed (absolute difference in velocity from the disk 
gas there) of HVCs, i.e., $\sim 240$~\kms\ \citep{wakker2004},
the kinetic energy and momentum 
would be $4.7\times 10^{51}\,\dtwenty^2$~ergs and 
$2.0\times 10^{6}\,\dtwenty^2$~\msol\kms\
including the He abundance, respectively.
This implies that the mass of CHVC040 
when it collided with the disk 
should have been an order of magnitude greater and that  
the HVC that we see (CHVC040) might be the remains of the original HVC.

The spatial morphology and velocity structure of CHVC040 
suggest that it is moving southwest in the plane of the sky 
and is approaching us.
The steep southwestern boundary of the cloud might represent the region
compressed by the interaction with the ambient medium, 
whereas the diffuse envelope might
be the material stripped off from the cloud due to  
the interaction or by the ram pressure of the surrounding medium \citep{santillan1999, kwak2009}.
The small velocity width along the southwestern boundary 
and large velocity width beyond appear to be consistent with such speculation.
Recently, \citet{heitsch2016} suggested that, for a CHVC with head-tail structure, 
the inclination angle, i.e., the angle between the CHVC's trajectory and the line-of-sight, can be derived from the asymmetry 
in position-velocity diagram along the head-tail line crossing the center of mass. 
The morphology of CHVC040 is different from typical head-tail CHVCs, i.e., CHVC040 has a wide flaring `tail' not a narrow elongated tail 
(see also \S~2), and its core is fragmented, so that it is not obvious if their model can be applied. 
Nevertheless if we assume that the center of mass is near the southwestern boundary 
and apply their model (their equations 1--4) 
to the position-velocity diagram along the red-dashed line in Figure 1, 
we obtain an inclination angle of $\sim 30\arcdeg$ which is consistent with our expectation.

The shape of CHVC040 appears to be pointed away from the 
Galactic midplane as  if it already went through the midplane. 
But the geometrical center of the supershell coincides with the current location 
of CHVC040, not being located in the midplane. 
Perhaps CHVC040 is approaching at an angle to the warped Galactic disk and has not yet fully penetrated the disk to 
the midplane. It is worth to note that, 
according to the \citet{jin2010}'s model, the GCN HVC stream is colliding to the Galactic plane almost perpendicularly {\em from below}, 
while the head-tail directions of individual HVCs seem to indicate that there is no preferential direction in their motions \citep{winkel2011}. 
Therefore, the orbit of CHVC040 is uncertain. 
The colliding geometry and the origin of the complex structures such as the hub and spokes in GS040 need to be explored.

\subsection{CHVC040 and Disruption of HVCs}

There are about three hundred known CHVCs in the Milky Way 
\citep{deheij2002, putman2002}. 
A considerable fraction of CHVCs has a head-tail structure 
indicating a ram pressure interaction with the diffuse galactic halo gas \citep{putman2011}. 
An important question is 
whether they are totally dissipated in the Galactic halo 
to feed the multi-phase circumgalactic medium
or they can survive their trip through the halo \citep[e.g.,][]{putman2011}. 
Since CHVC040 is located in the far outer Galaxy,
it may be of extragalactic origin rather than 
originating from a Galactic fountain, 
although it is not clear whether CHVC040 was originally 
a fragment of a nearby tidally disrupted galaxy 
or a cold cloud formed in a larger accreting flow 
of ionized, low-metallicity intergalactic gas.
Our result then directly shows that at least some CHVCs of extragalactic origin 
do survive and collide with the Galactic disk. 
According to numerical studies, 
CHVCs with \schi\ masses $\la 3\times 10^4$~\msol\ 
would be totally disrupted in the Galactic halo 
unless they are embedded in dark matter \citep{heitsch2009, plockinger2012}.
But dynamical shielding by an extended diffuse gaseous component can 
significantly extend their lifetime \citep{putman2012}. 
We have checked whether there are additional sources like the CHVC040-GS040 system
in the I-GALFA \schi\ data using the HVC catalog of \cite{deheij2002}.
There are twelve CHVCs in their Table~2 
including CHVC040 in the I-GALFA survey area, most of which are at relatively high
latitudes. They are isolated HVCs and have $\vlsr < -150$~\kms, so they 
belong to the GCN HVC complex. 
We see low-velocity \schi\ features around some CHVCs, but
none of them appear associated. 
A systematic study against all CHVCs may reveal 
other CHVC-supershell systems.

\acknowledgments

This work was supported by the National Research Foundation of Korea (NRF) grant 
funded by the Korea Government (MSIP) (No. 2012R1A4A1028713).



\vspace{5mm}

\begin{thebibliography}{}

\bibitem[Anderson et al.(2011)]{anderson2011} Anderson, L.~D., Bania, T.~M., Balser, D.~S., \& Rood, R.~T.\ 2011, \apjs, 194, 32
\bibitem[Anderson et al.(2014)]{anderson2014} Anderson, L.~D., Bania, T.~M., Balser, D.~S., et al.\ 2014, \apjs, 212, 1
\bibitem[Bagetakos et al.(2011)]{bagetakos2011} Bagetakos, I., Brinks, E., Walter, F., et al.\ 2011, \aj, 141, 23 
\bibitem[Braun \& Burton(1999)]{braun1999} Braun, R., \& Burton, W.~B.\ 1999, \aap, 341, 437
\bibitem[Br{\"u}ns et al.(2001)]{bruns2001} Br{\"u}ns, C., Kerp, J., \& Pagels, A.\ 2001, \aap, 370, L26 
\bibitem[Dame \& Thaddeus(2011)]{dame2011} Dame, T.~M., \& Thaddeus, P.\ 2011, \apjl, 734, L24 
\bibitem[de Heij et al.(2002)]{deheij2002} de Heij, V., Braun, R., \& Burton, W.~B.\ 2002, \aap, 391, 159 
\bibitem[Dennison et al.(1998)]{dennison1998} Dennison, B., Simonetti, J.~H., \& Topasna, G.~A.\ 1998, \pasa, 15, 147 
\bibitem[Draine(2011)]{draine2011} Draine, B.~T.\ 2011, {\it Physics of the Interstellar and Intergalactic Medium, ed. B. T. Draine} (Princeton Univ. Press)
\bibitem[Faridani et al.(2014)]{faridani2014} Faridani, S., Fl{\"o}er, L., Kerp, J., \& Westmeier, T.\ 2014, \aap, 563, A99 
\bibitem[Finkbeiner(2003)]{finkbeiner2003} Finkbeiner, D.~P.\ 2003, \apjs, 146, 407 
\bibitem[Gibson et al.(2012)]{gibson2012} Gibson, S.~J., Koo, B., Douglas, K.~A., et al.\ 2012, American Astronomical Society Meeting Abstracts, 219, \#349.29
\bibitem[Haffner et al.(2003)]{haffner2003} Haffner, L.~M., Reynolds, R.~J., Tufte, S.~L., et al.\ 2003, \apjs, 149, 405
\bibitem[Heiles(1979)]{heiles1979} Heiles, C.\ 1979, \apj, 229, 533
\bibitem[Heiles(1984)]{heiles1984} Heiles, C.\ 1984, \apjs, 55, 585 
\bibitem[Heitsch \& Putman(2009)]{heitsch2009} Heitsch, F., \& Putman, M.~E.\ 2009, \apj, 698, 1485 

\bibitem[Heitsch et al.(2016)]{heitsch2016} Heitsch, F., Bartell, B., Clark, S.~E., et al.\ 2016, arXiv:1606.06689 
\bibitem[Jin(2010)]{jin2010} Jin, S.\ 2010, \mnras, 408, L85 
\bibitem[Kang \& Koo(2007)]{kang2007} Kang, J.-h., \& Koo, B.-C.\ 2007, \apjs, 173, 85
\bibitem[Kamphuis et al.(1991)]{kamphuis1991} Kamphuis, J., Sancisi, R., \& van der Hulst, T.\ 1991, \aap, 244, L29 
\bibitem[Koo et al.(2010)]{koo2010} Koo, B.-C., Heiles, C., Stanimirovi{\'c}, S., \& Troland, T.\ 2010, \aj, 140, 262 
\bibitem[Kwak et al.(2009)]{kwak2009} Kwak, K., Shelton, R.~L., \& Raley, E.~A.\ 2009, \apj, 699, 1775 
\bibitem[Levine et al.(2006)]{levine2006} Levine, E.~S., Blitz, L., \& Heiles, C.\ 2006, \apj, 643, 881 
\bibitem[McClure-Griffiths et al.(2002)]{mcclure2002} McClure-Griffiths, N.~M., Dickey, J.~M., Gaensler, B.~M., \& Green, A.~J.\ 2002, \apj, 578, 176 
\bibitem[Mirabel \& Morras(1990)]{mirabel1990} Mirabel, I.~F., \& Morras, R.\ 1990, \apj, 356, 130 
\bibitem[Pl{\"o}ckinger \& Hensler(2012)]{plockinger2012} Pl{\"o}ckinger, S., \& Hensler, G.\ 2012, \aap, 547, AA43
\bibitem[Putman et al.(2002)]{putman2002} Putman, M.~E., de Heij, V., Staveley-Smith, L., et al.\ 2002, \aj, 123, 873 
\bibitem[Putman et al.(2011)]{putman2011} Putman, M.~E., Saul, D.~R., \& Mets, E.\ 2011, \mnras, 418, 1575\bibitem[Putman et al.(2012)]{putman2012} Putman, M.~E., Peek, J.~E.~G., \& Joung, M.~R.\ 2012, \araa, 50, 491 
\bibitem[Rhode et al.(1999)]{rhode1999} Rhode, K.~L., Salzer, J.~J., Westpfahl, D.~J., \& Radice, L.~A.\ 1999, \aj, 118, 323 
\bibitem[Santill{\'a}n et al.(1999)]{santillan1999} Santill{\'a}n, A., Franco, J., Martos, M., \& Kim, J.\ 1999, \apj, 515, 657
\bibitem[Tamanaha(1997)]{tamanaha1997} Tamanaha, C.~M.\ 1997, \apjs, 109, 139 
\bibitem[Tenorio-Tagle(1980)]{tenorio1980} Tenorio-Tagle, G.\ 1980, \aap, 88, 61 
\bibitem[Tenorio-Tagle et al.(1987)]{tenorio1987} Tenorio-Tagle, G., Franco, J., Bodenheimer, P., \& Rozyczka, M.\ 1987, \aap, 179, 219 
\bibitem[Voges et al.(1999)]{voges1999} Voges, W., Aschenbach, B., Boller, T., et al.\ 1999, \aap, 349, 389 
\bibitem[Wakker(1991)]{wakker1991a} Wakker, B.~P.\ 1991, \aap, 250, 499 
\bibitem[Wakker \& van Woerden(1991)]{wakker1991b} Wakker, B.~P., \& van Woerden, H.\ 1991, \aap, 250, 509 
\bibitem[Wakker(2004)]{wakker2004} Wakker, B.~P.\ 2004, \apss, 289, 381 
\bibitem[Westmeier et al.(2005)]{westmeier2005} Westmeier, T., Br{\"u}ns, C., \& Kerp, J.\ 2005, \aap, 432, 937 
\bibitem[Winkel et al.(2011)]{winkel2011} Winkel, B., Ben Bekhti, N., Darmst{\"a}dter, V., et al.\ 2011, \aap, 533, A105

\end{thebibliography}
\end{document}